\documentstyle[amssymb,prb,preprint,aps]{revtex}

\draft

\begin{document}
\title{Specific heat at low temperatures and magnetic measurements in Nd$_{0.5}$Sr$%
_{0.5}$MnO$_{3}$ and R$_{0.5}$Ca$_{0.5}$MnO$_{3}$ (R=Nd, Sm, Dy and Ho)
samples}
\author{J. L\'{o}pez and O. F. de Lima}
\address{Instituto de F\'{i}sica Gleb Wataghin, Universidade Estadual de Campinas,\\
UNICAMP, 13083-970, Campinas, SP, Brazil}
\author{P. N. Lisboa-Filho and F. M. Araujo-Moreira}
\address{Depto. de F\'{i}sica, Universidade Federal de S\~{a}o Carlos, CP-676,\\
S\~{a}o Carlos, SP, 13565-905, Brazil}
\maketitle

\begin{abstract}
We studied the magnetization as a function of temperature and magnetic field
in the compounds Nd$_{0.5}$Sr$_{0.5}$MnO$_{3}$, Nd$_{0.5}$Ca$_{0.5}$MnO$_{3}$%
, Sm$_{0.5}$Ca$_{0.5}$MnO$_{3}$, Dy$_{0.5}$Ca$_{0.5}$MnO$_{3}$ and Ho$_{0.5}$%
Ca$_{0.5}$MnO$_{3}$. Ferromagnetic, antiferromagnetic and charge ordering
transition in our samples agreed with previous reports. The intrinsic
magnetic moments of Nd$^{3+}$, Sm$^{3+}$, Dy$^{3+}$ and Ho$^{3+}$ ions
experienced a short range order at low temperatures. We also did specific
heat measurements with applied magnetic fields between 0 and 9 T and
temperatures between 2 and 30 K in all five samples. Below 10 K the specific
heat measurements evidenced a Schottky-like anomaly for all samples.
However, we could not successfully fit the curves to either a two level nor
a distribution of two-level Schottky anomaly.
\end{abstract}


\section{Introduction}

Compounds like La$_{0.5}$Ca$_{0.5}$MnO$_{3}$\ and Nd$_{0.5}$Sr$_{0.5}$MnO$%
_{3}$\ present a real-space ordering of Mn$^{3+}$ and Mn$^{4+}$ ions, named
as charge ordering (CO). Close to the charge ordering temperature (T$_{CO}$%
), these materials show anomalies in resistivity, magnetization and lattice
parameters as a function of temperature, magnetic field and isotope mass\cite
{Radaelli}$^{,}$\cite{Zhao2}. At low temperatures both, ferromagnetic and
antiferromagnetic phases, could coexist \cite{Moritomo}. However, a
relatively small external magnetic field can destroy the CO phase and
enforces a ferromagnetic orientation of the spins\cite{Xiao}.

Moreover, electron microscope analysis has revealed convincing evidence that
CO is accompanied by orientational ordering of the 3d$^{3}$ orbitals on the
Mn$^{3+}$ ions, called orbital ordering (OO)\cite{Mori}. The physical
properties in CO manganese perovskites are thought to arise from the strong
competition among a ferromagnetic double exchange interaction, an
antiferromagnetic superexchange interaction, and the spin-phonon coupling.
These interactions are determined by intrinsic parameters such as doping
level, average cationic size, cationic disorder and oxygen stoichiometry.
Microscopically, CO compounds are particularly interesting because spin,
charge and orbital degrees of freedom are at play simultaneously and
classical simplifications that neglect some of these interactions do not
work. More detailed information on the physics of manganites can be found in
a review paper by Myron B. Salamon and Marcelo Jaime\cite{Myron}.

We have shown that polycrystalline samples of La$_{0.5}$Ca$_{0.5}$MnO$_{3}$
and Nd$_{0.5}$Sr$_{0.5}$MnO$_{3}$ presented an unusual magnetic relaxation
behavior close to each critical temperature\cite{J.López1}, \cite{J.López2}.
However, a clear understanding of all these features has not been reached
yet. An alternative to a bulk characterization like magnetization would be
to perform specific heat measurements. In contrast to magnetization, which
has a vector character, the specific heat is an scalar. Therefore, a
comparison between both types of data could give valuable information.

L. Ghivelder et al.\cite{Ghivelder} reported specific heat measurements in
LaMnO$_{3+\delta }$ samples and they found that the specific heat at low
temperature is very sensitive to small variations of $\delta $, similarly to
results found by W. Schnelle et al.\cite{Schnelle} in a Nd$_{0.67}$Sr$%
_{0.33} $MnO$_{3-\delta }$ sample. In this latter work and also in a work by
J. E. Gordon et al.\cite{Gordon} a Schottky-like anomaly was found at low
temperatures in samples with similar compositions. They associated this
result to the magnetic ordering of Nd$^{3+}$ ions and to the crystal-field
splitting at low temperatures. F. Bartolom\'{e} et al.\cite{Bartolomé} also
found a Schottky-like anomaly in a closely related compound of NdCrO$_{3}$.
They proposed a crystal-field energy level scheme in agreement with
neutron-scattering studies in the same sample. In two papers V. N.
Smolyaninova et al.\cite{Smolyaninova1}, \cite{Smolyaninova2} studied the
low temperature specific heat in Pr$_{1-x}$Ca$_{x}$MnO$_{3}$ (0.3%
\mbox{$<$}%
x%
\mbox{$<$}%
0.5) and La$_{1-x}$Ca$_{x}$MnO$_{3}$ (x=0.47, 0.5 and 0.53). They found an
excess specific heat, C%
\'{}%
(T), of non-magnetic origin associated with charge ordering. They also
reported that a magnetic field sufficiently high to induce a transition from
the charge ordered state to the ferromagnetic metallic state did not
completely remove C%
\'{}%
(T). However, no Schottky anomaly was found in any of these compounds.

Here, we report a general magnetic characterization and specific heat
measurements with applied magnetic fields between 0 and 9 T and temperatures
between 2 and 30 K for Nd$_{0.5}$Sr$_{0.5}$MnO$_{3}$, Nd$_{0.5}$Ca$_{0.5}$MnO%
$_{3}$, Sm$_{0.5}$Ca$_{0.5}$MnO$_{3}$, Dy$_{0.5}$Ca$_{0.5}$MnO$_{3}$ and Ho$%
_{0.5}$Ca$_{0.5}$MnO$_{3}$ samples. All these compounds presented a
Schottky-like anomaly at low temperatures. We have already reported a short
version of preliminary results about these topics \cite{JLópez3}. However,
as far as we know, detailed specific heat measurements in these compounds
have not been published yet.

\section{Experimental methods}

Polycrystalline samples of Nd$_{0.5}$Sr$_{0.5}$MnO$_{3}$, Nd$_{0.5}$Ca$%
_{0.5} $MnO$_{3}$ and Ho$_{0.5}$Ca$_{0.5}$MnO$_{3}$ were prepared by the
sol-gel method\cite{Paulo}. Stoichiometric parts of Nd$_{2}$O$_{3}$ (Ho$_{2}$%
O$_{3}$) and MnCO$_{3}$ were dissolved in HNO$_{3}$ and mixed to an aqueous
citric acid solution, to which SrCO$_{3}$ or CaCO$_{3}$\ was added. The
mixed metallic citrate solution presented the ratio citric acid/metal of 1/3
(in molar basis). Ethylene glycol was added to this solution, to obtain a
citric acid/ethylene glycol ratio 60/40 (mass ratio). The resulting solution
was neutralized to pH$\sim $7 with ethylenediamine. This solution was turned
into a gel, and subsequently decomposed to a solid by heating at 400 $^{o}$%
C. The resulting powder was heat-treated in vacuum at 900 $^{o}$C for 24
hours, with several intermediary grindings, in order to prevent formation of
impurity phases. This powder was pressed into pellets and sintered in air at
1050 $^{o}$C for 12 hours.

Polycrystalline samples of Sm$_{0.5}$Ca$_{0.5}$MnO$_{3}$ and Dy$_{0.5}$Ca$%
_{0.5}$MnO$_{3}$ were prepared from stoichiometric amounts of Sm$_{2}$O$_{3}$
or Dy$_{2}$O$_{3}$, CaO, and MnO$_{2}$ by standard solid-state reaction
method. Purity of these starting materials was more than 99.99 \%. All the
powders were mixed and grinded for a long time in order to produce a
homogeneous mixture. First, the mixture was heated at 927 $^{o}$C for 24
hours and after that it was grinded and heated at 1327 $^{o}$C (72 hours)
and 1527 $^{o}$C (48 hours). X-ray diffraction measurements indicated high
quality samples in all five cases.

The magnetization measurements were done with a Quantum Design MPMS-5S SQUID
magnetometer. Specific heat measurements were done with a Quantum Design
PPMS calorimeter. The PPMS used the two relaxation time technique, and data
was always collected during sample cooling. The intensity of the heat pulse
was calculated to produce a variation in the temperature bath between 0.5 \%
(at low temperatures) and 2\% (at high temperatures). Experimental errors
during the specific heat and magnetization measurements were lower than 1 \%
for all temperatures and samples.

\section{Results and Discussion}

\subsection{Magnetization measurements}

Figure 1 shows the temperature dependence of magnetization, measured with a
5 T applied magnetic field and zero field cooling conditions, in
polycrystalline samples of Nd$_{0.5}$Sr$_{0.5}$MnO$_{3}$, Nd$_{0.5}$Ca$%
_{0.5} $MnO$_{3}$, Sm$_{0.5}$Ca$_{0.5}$MnO$_{3}$, Dy$_{0.5}$Ca$_{0.5}$MnO$%
_{3}$ and Ho$_{0.5}$Ca$_{0.5}$MnO$_{3}$. The curves are plotted with a
logarithmic scale in the y-axes to allow the comparison of the samples.
Charge ordering transition temperatures (T$_{CO}$) are indicated by arrows
at 160, 250, 270, 280 and 271 K, respectively. These temperatures are
associated to peaks in the magnetization curves, in agreement with previous
reports \cite{Kajimoto}$^{,}$ \cite{Millange}$^{,}$\cite{Tokura}$^{,}$\cite
{Terai}. It is interesting to note that the relation between the charge
ordering temperature and the antiferromagnetic ordering temperature (T$_{N}$%
) changes from one sample to the other\cite{Kajimoto}$^{,}$ \cite{Millange}$%
^{,}$\cite{Tokura}$^{,}$\cite{Terai}. In the first case they are
approximately coincident, in the second and third cases the charge ordering
temperatures are much higher, and in the fourth and fifth cases a long range
antiferromagnetic transition is not observed.

The Nd$_{0.5}$Sr$_{0.5}$MnO$_{3}$ sample presented a ferromagnetic
transition at T$_{C}\thickapprox $250 K and an antiferromagnetic transition
at T$_{N}\thickapprox $160 K. The Nd$_{0.5}$Ca$_{0.5}$MnO$_{3}$ compound
presented a strong maximum near T$_{CO}$, but showed an unexpected minimum
close to the antiferromagnetic transition temperature T$_{N}\thickapprox $%
160 K. Usually an antiferromagnetic transition is accompanied by a maximum
in the temperature dependence of the magnetization. For temperatures lower
than 10, 20 and 50 K the Nd$_{0.5}$Sr$_{0.5}$MnO$_{3}$, Nd$_{0.5}$Ca$_{0.5}$%
MnO$_{3}$ and Sm$_{0.5}$Ca$_{0.5}$MnO$_{3}$ samples showed a sharp increase
in the magnetization. This trend have been associated to a short range
magnetic ordering of the intrinsic magnetic moment of Nd$^{3+}$ ions\cite
{Mathieu}. However, no long range ferromagnetic order of the Nd$^{3+}$ ions
was found in neutron diffraction measurements at these low temperatures\cite
{Kajimoto}$^{,}$ \cite{Millange}.

Differently from the three previous samples, the Dy$_{0.5}$Ca$_{0.5}$MnO$%
_{3} $ and Ho$_{0.5}$Ca$_{0.5}$MnO$_{3}$ compounds do not present a strong
maximum at the charge ordering temperature in the magnetization versus
temperature curve. However, a clear inflection is observed at T$_{CO}$ for
both samples, as revealed by the temperature derivative shown in the inset
of fig. 1b. The existence of charge ordering in Dy$_{0.5}$Ca$_{0.5}$MnO$_{3}$
and Ho$_{0.5}$Ca$_{0.5}$MnO$_{3}$ was suggested by\ T. Terai et al. \cite
{Terai} after studies of magnetization and resistivity curves. Our high
temperature measurements of specific heat (not shown) presented peaks at
around the same temperature interval of the suggested charge ordered
transition. These results will be published elsewhere.

Figure 2 shows the magnetization hysteresis loops at 2 K for the five
studied samples. The applied magnetic field was increased from 0 to 5 T,
decreased to -5 T and then increased back to 5 T again. The Nd$_{0.5}$Sr$%
_{0.5}$MnO$_{3}$\ curve is characteristic of a two phases mixture: one
ferromagnetic and another antiferromagnetic. The ferromagnetic part is
easily oriented at low magnetic field values and shows a hysteretic
behavior. The almost linear and reversible dependence for magnetic fields
higher than approximately 1 T, indicates a gradual destruction of the
antiferromagnetic phase\cite{J.López2}. R. Mahendiram et al.\cite{Mahendiran}
reported, in a sample with the same composition, and measured at 50 K, that
for magnetic fields higher than approximately 5 T the magnetization started
to increase rapidly, and for magnetic fields above 10 T, it slowly
approached the ferromagnetic saturation value. Because at 2 K these
transition fields are expected to be much higher than 5 T we were unable to
see them.

The Nd$_{0.5}$Ca$_{0.5}$MnO$_{3}$\ and Sm$_{0.5}$Ca$_{0.5}$MnO$_{3}$ curves
only showed a small trace of ferromagnetic component at very low fields.
Beyond this region the curves are linear and reversible for the whole
magnetic field interval. As before, this linearity characterizes the gradual
destruction of the antiferromagnetic phase. F. Millange et al.\cite{Millange}
reported for a Nd$_{0.5}$Ca$_{0.5}$MnO$_{3}$\ sample a set of M vs. H curves
with applied magnetic fields up to 22 T. They found sharp transitions and a
large hysteresis at 130 K, between applied magnetic fields of 12 and 18 T.
Their results were interpreted as evidence of the existence of a spin-flop
transition. Y. Tokura et al.\cite{Tokura} reported the phase diagram (H vs.
T) for several charge ordered compounds and found that the Sm$_{0.5}$Ca$%
_{0.5}$MnO$_{3}$ sample required the largest magnetic fields to destroy the
charge ordering state. For example, at about 10 K the transition fields
were\ approximately 39 T and 65 T. In the M vs. H curves for the Dy$_{0.5}$Ca%
$_{0.5}$MnO$_{3}$ and Ho$_{0.5}$Ca$_{0.5}$MnO$_{3}$ samples, shown in fig.
2b, two characteristics are well noticed. One, that there is not hysteresis
at all, and the other, that the magnetization values at 5 T are well below
the theoretical saturation values of 8.73 and 8.70 $\mu _{B}$, respectively 
\cite{Millange}.

\subsection{Specific heat at low temperatures}

Figure 3 reproduces the specific heat measurements for temperatures between
2 and 30 K in the samples of (a) Nd$_{0.5}$Sr$_{0.5}$MnO$_{3}$, (b) Nd$%
_{0.5} $Ca$_{0.5}$MnO$_{3}$ (c) Sm$_{0.5}$Ca$_{0.5}$MnO$_{3}$, (d) Dy$_{0.5}$%
Ca$_{0.5}$MnO$_{3}$ and (e) Ho$_{0.5}$Ca$_{0.5}$MnO$_{3}$. Measurements were
made in the presence of applied magnetic fields of 0 T (squares), 5 T
(circles), 7 T (up triangles) and 9 T (down triangles). Note that close to 5
K all curves show a Schottky-like anomaly\cite{Kittel}.

It is important to stress that high values of specific heat were found at
low temperatures for the five samples presented here. These values are
similar to those reported by J. E. Gordon et. al.\cite{Gordon} in a sample
of Nd$_{0.67}$Sr$_{0.33}$MnO$_{3}$. However, the absolute values of specific
heat at 2 K, reported by J. J. Hamilton et al.\cite{Hamilton} in samples of
La$_{0.67}$Ba$_{0.33}$MnO$_{3}$ and La$_{0.80}$Ca$_{0.20}$MnO$_{3}$, were
more than 100 times smaller. Similarly, V. Hardy et al.\cite{Hardy}, in a
single crystal of Pr$_{0.63}$Ca$_{0.37}$MnO$_{3}$, found values
approximately equal to those reported by J. J. Hamilton et al.\cite{Hamilton}%
. The high values of specific heat found in our work indicates enhanced
excitations and could be due to an increase in the effective mass of the
electrons caused by localization. This interpretation is consistent with the
insulating behavior revealed by electrical resistivity measurements\cite
{Kajimoto}$^{,}$ \cite{Millange}$^{,}$ \cite{Tokura}$^{,}$ \cite{Terai}.The
fact that resistivity is increasing with decreasing temperatures is
associated with localization of charge carriers, which lead to the increase
of their effective mass.

Continuous lines in figure 3 indicate the fitting of the experimental data
between 15 and 30 K by the following expression\cite{Gordon}:

\begin{equation}
C=\sum \beta _{2n+1}\,T^{\text{\/}2n+1}  \label{5}
\end{equation}
Here, ${\em C}$ is the specific heat, ${\em T}$ is the temperature and the
parameters $\beta _{2n+1}$ represent the contribution of phonon modes.
Notice that we did not include the lowest temperature interval to avoid the
Schottky anomaly. To be able to fit the whole temperature interval we have
chosen values of $n$ from 1 to 4. Nonetheless, this large number of\ free
parameters turns difficult an unique determination of each one.

Since from resistivity measurements\cite{Kawano}$^{,}$ \cite{Millange}$^{,}$ 
\cite{Terai} all the studied samples show an insulating behavior at low
temperatures, and the applied magnetic fields are not strong enough to
destroy this characteristic, one should not expect the linear contribution
from the free electrons to the specific heat. However, other kind of
excitations could also lead to a linear contribution. This can imply an
implicit error of the fitting model. Moreover, we could not resolve in our
data a term of type T$^{3/2}$, usually interpreted as an evidence to the
occurrence of ferromagnetic interactions. However, previous studies in
samples clearly identified by other techniques as ferromagnetic, have not
found this term in the specific heat either\cite{Smolyaninova2}$^{,}$ \cite
{Hamilton}. We also tried to include the hyperfine contribution\ with a T$%
^{-2}$ dependence, but the fitting did not improve. Gordon et al.\cite
{Gordon} fitted the hyperfine contribution below 1.5 K, a temperature
interval that we are not able to measure at the moment.

The influences of the external magnetic field on the specific heat are not
clear for all samples. Almost no dependence was found for the Sm$_{0.5}$Ca$%
_{0.5}$MnO$_{3}$ sample, while in the other cases the specific heat
increases with the external magnetic field. Even though magnetic
interactions do exist between 15 and 30 K, and considering that there is not
a long range magnetic phase transition in this interval, we believe that the
relative contribution of the magnetic excitations in this temperature
interval, compared with the lattice vibrations, should be small. The
external magnetic field could be primarily affecting the lattice vibrations,
due to lattice distortions induced by the field, rather than affecting the
antiferromagnetic or ferromagnetic spins waves. As a comparison, recently A.
N. Lavrov et al.\cite{Lavrov} described an unexpected magnetic effect on
crystal shape, in which the direction of the crystal's axes were swapped\
and the shape changed when a magnetic field was applied; this in turn
induced curious memory effects in resistivity and magnetic susceptibility of
an antiferromagnets, La$_{2-x}$Sr$_{x}$CuO$_{4}$.

The values of $\beta _{3}$ change between 0.28 mJ/(mol K$^{4}$) at H=0 T in
Nd$_{0.5}$Sr$_{0.5}$MnO$_{3}$ to 1.57 mJ/(mol K$^{4}$) at H=5 T in Ho$_{0.5}$%
Ca$_{0.5}$MnO$_{3}$. The corresponding Debye temperatures (T$_{D}$),
obtained from $\beta _{3}$, are plotted in figure 4a. The graphs in the
insets of figure 3 show the differences between the specific heat
experimental data and the phonon contribution to the specific heat,
extrapolated to low temperatures from the fitting in the temperature
interval between 15 and 30 K.

Figure 4 shows (a) the magnetic field dependence of the Debye temperature,
(b) the variation of magnetic entropy between 2 and 20 K ($\Delta S$) and
(c) the Schottky temperature (T$_{S}$) in the five studied samples. The
Debye temperature was calculated using the values of $\beta _{3}$\ and the
following equation\cite{Kittel}:

\begin{equation}
T_{D}=\left( \frac{12\pi ^{4}nR}{5\beta _{3}}\right) ^{1/3}  \label{6}
\end{equation}
where ${\em n}$ is the number of atoms in the unit cell and ${\em R}$ is the
ideal gas constant. We should point out that the Debye temperature was
estimated from the low temperature data. This procedure leads to values that
are slightly different than the actual values of {\em T}$_{D}$ for which the
specific heat saturates. As shown in fig.4a our {\em T}$_{D}$ values in
general decrease with the increase of the applied magnetic field. We have
also made specific heat measurements with a 9 T magnetic field, at high
temperatures, for several charge ordered compounds\cite{JLópez4}, and they
are in agreement with the magnetic field dependence of the Debye
temperatures shown here. Other authors\cite{Gordon} have made an initial
assumption that the Debye temperature is magnetic field independent, which
is not supported by our experimental results.

To calculate the variation in entropy ($\Delta $S), associated to the
Schottky anomaly, we used the definition:

\begin{equation}
\Delta S=\int\limits_{Ti}^{Tf}\frac{\left( C-C_{ph}\right) }{T}\,dT
\label{4}
\end{equation}
where ${\em T}_{i}$ and ${\em T}_{f}$ are two temperatures conveniently
chosen to delimitate the interval of interest and ${\em C}_{ph}$ is the
specific heat due to the lattice oscillations.

The entropy variation, associated to the Schottky anomaly, grows as a
function of magnetic field in the samples of Nd$_{0.5}$Sr$_{0.5}$MnO$_{3}$
and Sm$_{0.5}$Ca$_{0.5}$MnO$_{3}$. The same result is clearly visualized
from the height of the Schottky anomaly in the insets of figures 3a e 3c. J.
E. Gordon et al.\cite{Gordon} also reported a similar increase in a sample
of Nd$_{0.67}$Sr$_{0.33}$MnO$_{3}$. However, the Schottky entropy variation
decreases with the increase of magnetic field in the Nd$_{0.5}$Ca$_{0.5}$MnO$%
_{3}$, Dy$_{0.5}$Ca$_{0.5}$MnO$_{3}$ and Ho$_{0.5}$Ca$_{0.5}$MnO$_{3}$
samples. In these three cases, the C vs. T curves in figure 3b, 3d and 3e,
indicate that the local minimum, at a temperature above the Schottky
anomaly, disappears with the increase of the external magnetic field. This
is also reflected in the decrease of the height of the peak with the
increase of the applied magnetic field (insets of figures 3b, 3d and 3e).

The expected entropy variation from the magnetic ordering of Nd$^{3+}$, Sm$%
^{3+}$, Dy$^{3+}$ or Ho$^{3+}$ ions could be estimated\cite{Gordon} as $%
\Delta S$=0.5 R ln(2), where ${\em R}$ is the ideal gas constant. The
actually found variation correspond to values from 63 to 77 \% of the
expected ones in the Nd$_{0.5}$Sr$_{0.5}$MnO$_{3}$ sample, 80 to 62 \% in
the Nd$_{0.5}$Ca$_{0.5}$MnO$_{3}$ sample, 42 to 62 \% in the Sm$_{0.5}$Ca$%
_{0.5}$MnO$_{3}$ sample, 52 to 19 \% in the Dy$_{0.5}$Ca$_{0.5}$MnO$_{3}$
sample and 68 to 25 \% in the Ho$_{0.5}$Ca$_{0.5}$MnO$_{3}$ sample, for
magnetic fields between 0 and 9 T, respectively. J. E. Gordon et al.\cite
{Gordon} found that the entropy variation associated to the ordering of Nd$%
^{3+}$ ions\ in Nd$_{0.67}$Sr$_{0.33}$MnO$_{3}$ was approximately 85 \% of
the expected value.

Figure 4c shows that T$_{S}$, determined from the maxima\bigskip\ in the
insets of figure 3, grows with the increase of the external magnetic field
in most of the cases. T$_{S}$ in the Dy$_{0.5}$Ca$_{0.5}$MnO$_{3}$ sample
first increased and later slightly decreased. The growth of T$_{S}$ seems to
be saturated for a magnetic field of 5 T in the sample of Ho$_{0.5}$Ca$%
_{0.5} $MnO$_{3}$ and it is almost constant in the Sm$_{0.5}$Ca$_{0.5}$MnO$%
_{3}$ sample. It is also interesting to note here the relative low T$_{S}$
values. For the reagent compounds of Nd$_{2}$O$_{3}$, Sm$_{2}$O$_{3}$, Dy$%
_{2}$O$_{3} $ and Ho$_{2}$O$_{3}$ the peak in the specific heat, measured in
zero magnetic field, were found at approximately 10, 7, 7 and 9 K,
respectively \cite{Touloukian}. The fact that T$_{S}$\ is lower in the
manganese compounds suggests that the collective charge ordered phase could
be determining a smaller splitting in the energy levels. It also indicates
that an interpretation of the Schottky like anomaly based only on the ion
total angular momentum or degeneracy is not adequate and one should consider
the sample crystalline structure in detail.

Let us consider that Nd (Ho) ions be oriented by a molecular field
interaction (H$_{mf}$), and not by the exchange interaction between pairs of
Nd-Nd ions (Ho-Ho)\cite{Gordon}. Assuming that H$_{mf}$\ does not change
with the external magnetic field, it is possible to estimate it, using a
mean field model and the peak temperature in the specific heat. Considering
a two level energy splitting ${\em \Delta (H)}$, due to a magnetic moment $%
{\em m}${\em \ }in an external magnetic field ${\em H}$, one finds in zero
applied field $\Delta (0)=2mH_{mf}$, and in ${\em H=9}$ T the value changes
to $\Delta (9$ T$)=2m[H_{mf}+9$ T$]$. One can also use that the energy
splitting could be related to the peak temperature in the specific heat by $%
\Delta {\em \ =k}_{B}{\em \ T}_{S}{\em /0.418}$, a relation valid for a two
level Schottky function\cite{Kittel}. Solving this system of two linear
equations we found that ${\em H}_{mf}{\em \ =11.4}$ T and ${\em m=0.43\,}\mu
_{B}$ \ in Nd$_{0.5}$Sr$_{0.5}$MnO$_{3}$, ${\em H}_{mf}{\em \ =20.6}$ T and $%
{\em m=0.44\,}\mu _{B}$ in Nd$_{0.5}$Ca$_{0.5}$MnO$_{3}$,\ ${\em H}_{mf}{\em %
\ =18.3}$ T and ${\em m=0.42\,}\mu _{B}$ in Dy$_{0.5}$Ca$_{0.5}$MnO$_{3}$\
and ${\em H}_{mf}{\em \ =13.6}$ T and ${\em m=0.58\,}\mu _{B}$ in Ho$_{0.5}$%
Ca$_{0.5}$MnO$_{3}$.\ These values of ${\em m}$ are smaller than those
obtained from susceptibility measurements at high temperatures\cite{Millange}%
. However, they are similar to the ones found by J. E. Gordon et al.\cite
{Gordon} using the same method ($H_{mf}=$10 T and $m=$ 0.8 $\mu _{B}$ in Nd$%
_{0.67}$Sr$_{0.33}$MnO$_{3}$). This model could not be applied to the Sm$%
_{0.5}$Ca$_{0.5}$MnO$_{3}$ sample because T$_{S}$ did not change very much
with the applied magnetic field.

The ground state of the Nd$^{3+}$, Sm$^{3+}$, Dy$^{3+}$ and Ho$^{3+}$ ions
are usually denoted as $^{4}$I$_{9/2}$,\ $^{6}$H$_{5/2}$, $^{6}$H$_{15/2}$,
and $^{5}$I$_{8}$, where {\em I\ }or {\em H} stands for an orbital angular
momentum L=6 or L=5, the superprefix specify the total spin as ${\em 2S+1}$
and the subscript the total angular momentum ${\em J}$. The number of the
lowest energy levels is given by ${\em 2J+1}$, which leads to 5, 3 and 8
Kramers doublets in the ground state of the first, second and third ion,
respectively. The Ho$^{3+}$ ions has a singlet and 8 Kramers doublets\cite
{Ashcroft}. F. Bartolom\'{e} et. al.\cite{Bartolomé} showed that the second
doublet in the Nd$^{3+}$ ion was approximately 120 K (in energy) above the
lowest doublet. As this temperature is about 10 times higher than the
temperature where the Schottky anomaly appears, the contribution of the
second doublet is expected to be small. The second doublet is even higher in
temperature for the Dy$^{3+}$ and Ho$^{3+}$ ions and slightly lower for the
Sm$^{3+}$ ion in comparison with the Nd$^{3+}$ ion.

In a previous report\cite{JLópez3} we showed that a two level Schottky
function (only one doublet) did not fit properly our experimental data at
low temperatures. The same result was verified for all the new experimental
data presented here. One alternative, justified by the existence of several
different grains in polycrystalline samples, is to consider a distribution
of energy splitting around the value that would correspond to a single
crystal in the same two level Schottky model. Although the fitting results
using this second approach improved a little bit, we found that they still
remained unsatisfactory.

At first sight someone might be tempted to correlate the existence of the
Schottky anomaly with the presence of an intrinsic magnetic moment in Nd$%
^{3+}$ and Ho$^{3+}$ ions (in contrast with La$^{3+}$ ions without magnetic
moment and no Schottky anomaly in the manganite). However, specific heat
measurements reported by V. Hardy et al.\cite{Hardy} in a compound of Pr$%
_{0.63}$Ca$_{0.37}$MnO$_{3}$ (Pr$^{3+}$ ions have approximately the same
magnetic moment as Nd$^{3+}$ ions) did not show any Schottky anomaly.
Moreover, Ho$^{3+}$ ions have an intrinsic magnetic moment almost 3 times
bigger than Nd$^{3+}$ ions, but the Schottky temperature at zero magnetic
field were 2.73 K in Nd$_{0.5}$Sr$_{0.5}$MnO$_{3}$, 5.08 K in Nd$_{0.5}$Ca$%
_{0.5}$MnO$_{3}$, and 4.39 K in Ho$_{0.5}$Ca$_{0.5}$MnO$_{3}$.

Probably the existence of the Schottky anomaly is related with the Kramers
theorem. It states that an ion possessing an odd number of electrons, no
matter how unsymmetrical the crystal field, must have a ground state that is
at least doubly degenerate\cite{Ashcroft}. This could lead to the thermal
depopulation that produces the Schottky anomaly in the specific heat. Ions
of Ce, Nd, Sm, Gd, Dy, Er and Yb all have an odd number of electrons and
their respective oxides present a Schottky anomaly in the specific heat.
However, the Kramers theorem does not exclude that ions with an even number
of electrons might also have a doubly degenerate ground state. For instance,
someone might be tempted to state that Ho ions are equivalent to Pr ions and
therefore no Schottky like anomaly should be expected, however experiments
prove this analogy to be wrong.

The physics discussed here do not allow us to separate a single variable
like: degeneracy of the rare earth ions ground state, effective magnetic
moment of rare earth ions, or even talk of the rare earth ion instead of the
sample crystalline structure, to explain the observed results. All we can
say is that no single variable, considered individually, can explain the
experiments.

\section{Conclusions}

We have made a magnetic characterization of Nd$_{0.5}$Sr$_{0.5}$MnO$_{3}$, Nd%
$_{0.5}$Ca$_{0.5}$MnO$_{3}$, Sm$_{0.5}$Ca$_{0.5}$MnO$_{3}$, Dy$_{0.5}$Ca$%
_{0.5}$MnO$_{3}$ and Ho$_{0.5}$Ca$_{0.5}$MnO$_{3}$ polycrystalline samples.
Ferromagnetic, antiferromagnetic and charge ordering transitions in our
samples agreed with previous reports. We also reported, to our knowledge for
the first time, specific heat measurements with applied magnetic fields
between 0 and 9 T and temperatures between 2 and 30 K in all those five
samples. Absolute values of specific heat close to 2 K were about 100 times
higher in our samples than in other charge ordering samples like Pr$_{0.63}$%
Ca$_{0.37}$MnO$_{3}$. At low temperatures the specific heat curve, in all
five studied samples and measured magnetic fields, showed a Schottky-like
anomaly. In almost all cases an increase in the applied magnetic field moves
the Schottky peak to higher temperatures. However, the position of the peak
is almost independent of the applied magnetic field in the Sm$_{0.5}$Ca$%
_{0.5}$MnO$_{3}$\ sample. We could not successfully fit the curves by either
assuming a single or a distribution of two-level-Schottky anomaly. More
experiments are clearly necessary to unambiguously identify the origin of
the Schottky anomaly and its possible correlation with the charge ordered
phase.

We thank the Brazilian science agencies FAPESP and CNPq for the financial
support.

\bigskip

\newpage

\begin{center}
FIGURE CAPTIONS
\end{center}

Figure 1. Temperature dependence of the magnetization, with a 5 T applied
magnetic field, in field cooling--warming condition for the five
polycrystalline samples studied. Magnetization is given in Bohr magnetons
per manganese ion. The Curie (T$_{C}$), Ne\'{e}l (T$_{N}$) and charge
ordering (T$_{CO}$) temperatures are indicated for each curve. The curves
are plotted with a logarithmic scale in the y-axes to allow the comparison
of all samples. The inset in fig.1b represents the temperature derivative of
the magnetization close to the charge ordering transition.

Figure 2. Magnetization versus applied magnetic field at 2 K for the five
samples studied. After a zero field cooling the magnetic field was increased
from 0 to 5 T, decreased from 5 T to -5 T and increased again from -5 T to 5
T. Magnetization is given in Bohr magnetons per manganese ion.

Figure 3. Specific heat measurements between 2 and 30 K in the samples of
(a) Nd$_{0.5}$Sr$_{0.5}$MnO$_{3}$, (b) Nd$_{0.5}$Ca$_{0.5}$MnO$_{3}$, (c) Sm$%
_{0.5}$Ca$_{0.5}$MnO$_{3}$, (d) Dy$_{0.5}$Ca$_{0.5}$MnO$_{3}$\ and (e) Ho$%
_{0.5}$Ca$_{0.5}$MnO$_{3}$. Measurements were made in the presence of
applied magnetic fields of 0 T (squares), 5 T (circles), 7 T (up triangles)
and 9 T (down triangles). Continuous lines represent the fitting of the 15
to 30 K temperature interval data by the phonon contribution, as explained
in the text. The graphs in the insets show the difference between the
experimental values and the extrapolation of the phonon contribution to
temperatures lower than 15 K.

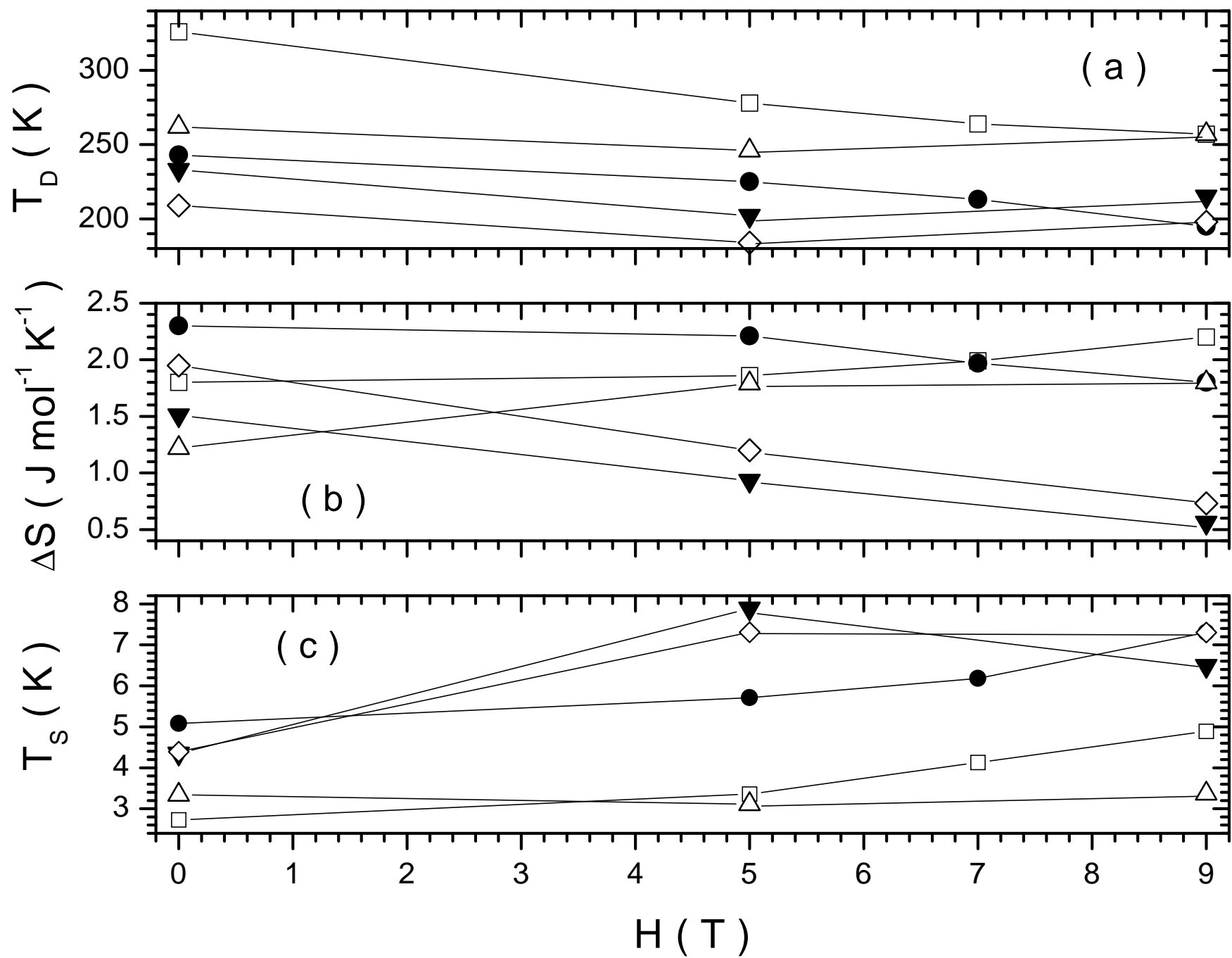
Figure 4. Debye temperature (T$_{D}$), entropy variation between 2 and 20 K (%
$\Delta S$) and Schottky temperature (T$_{S}$) as a function of the applied
magnetic field in Nd$_{0.5}$Sr$_{0.5}$MnO$_{3}$ (open squares), Nd$_{0.5}$Ca$%
_{0.5}$MnO$_{3}$ (closed circles), Sm$_{0.5}$Ca$_{0.5}$MnO$_{3}$ (open up
triangles), Dy$_{0.5}$Ca$_{0.5}$MnO$_{3}$ (closed down triangles) and Ho$%
_{0.5}$Ca$_{0.5}$MnO$_{3}$ (open diamond).

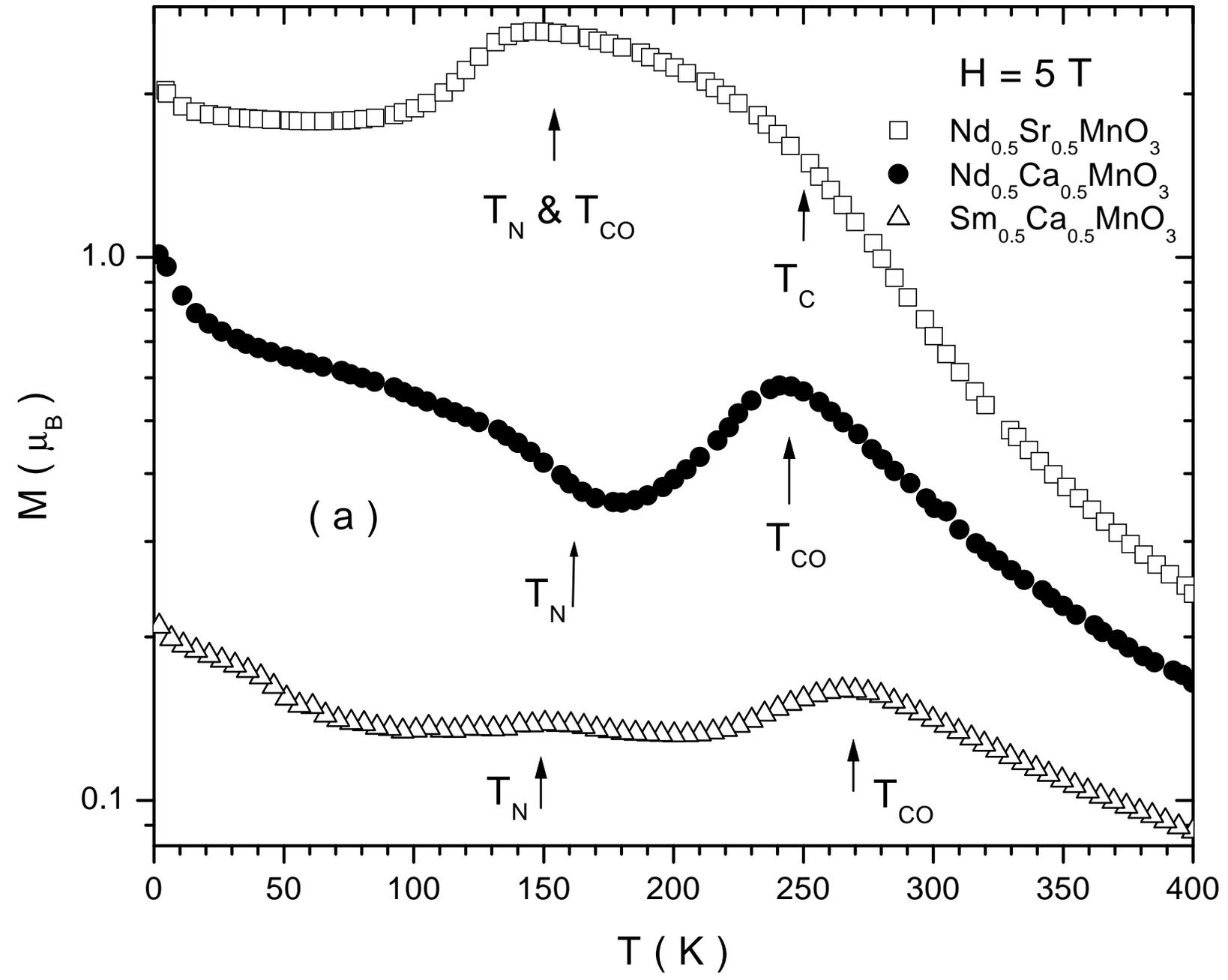
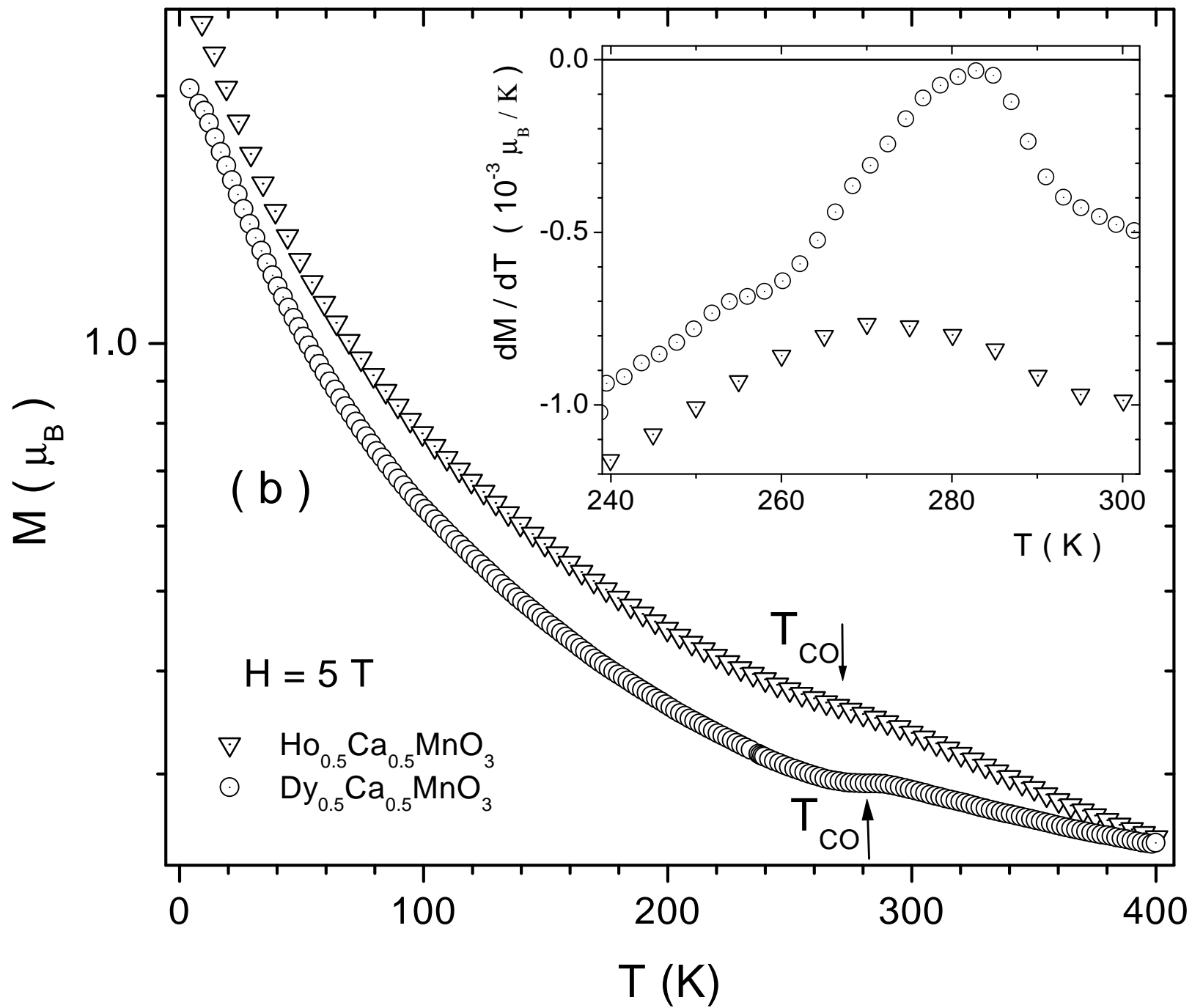
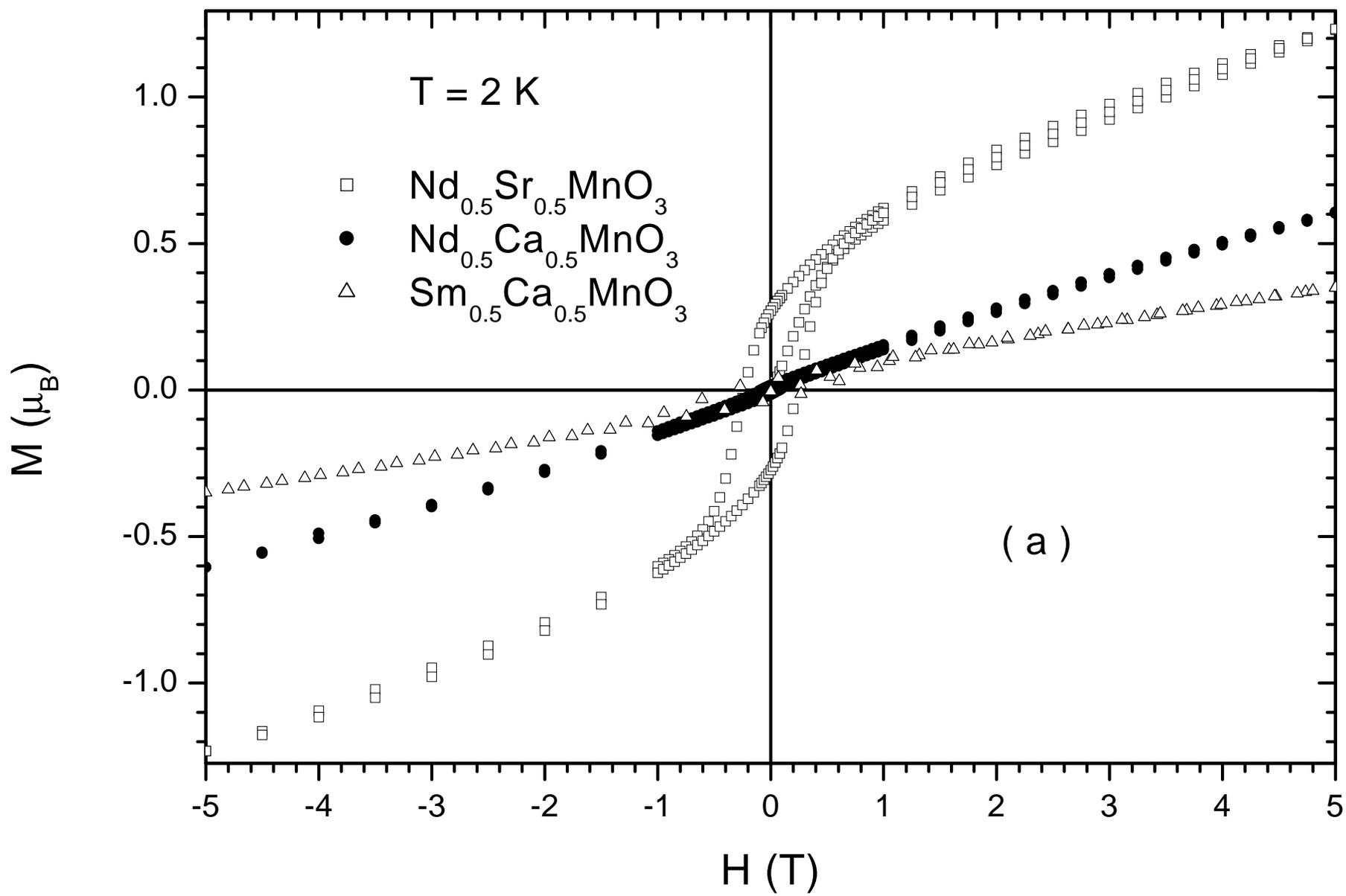
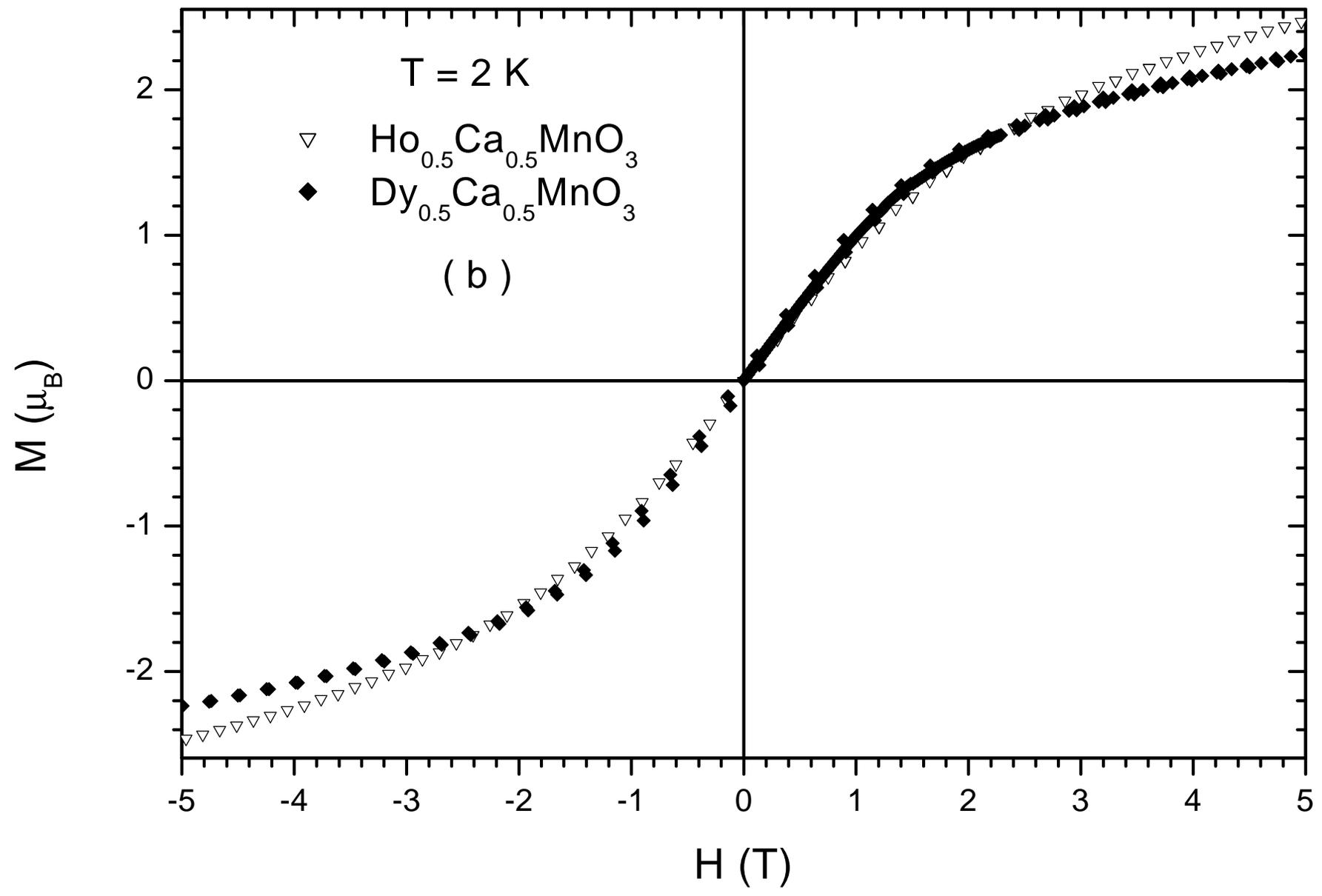
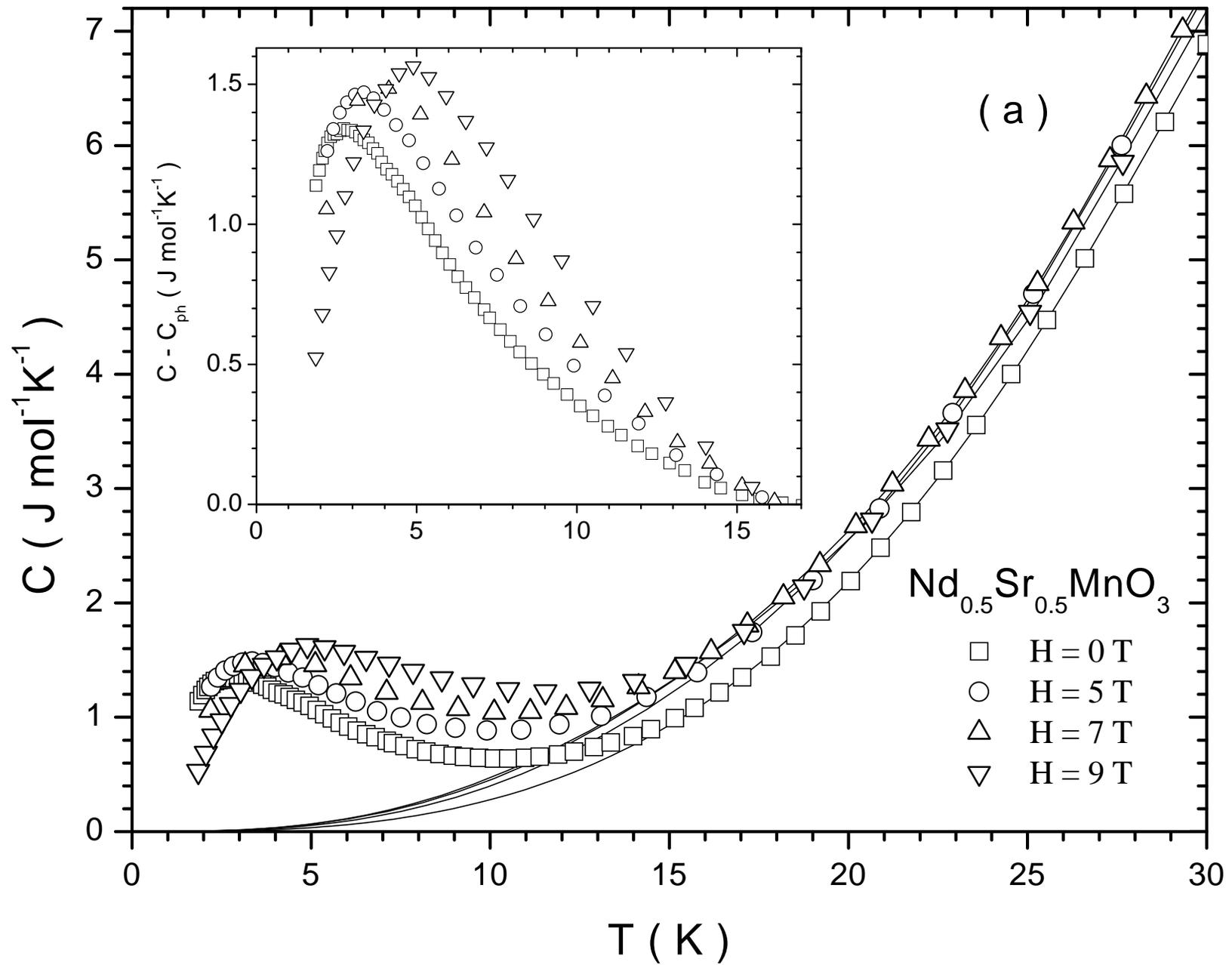
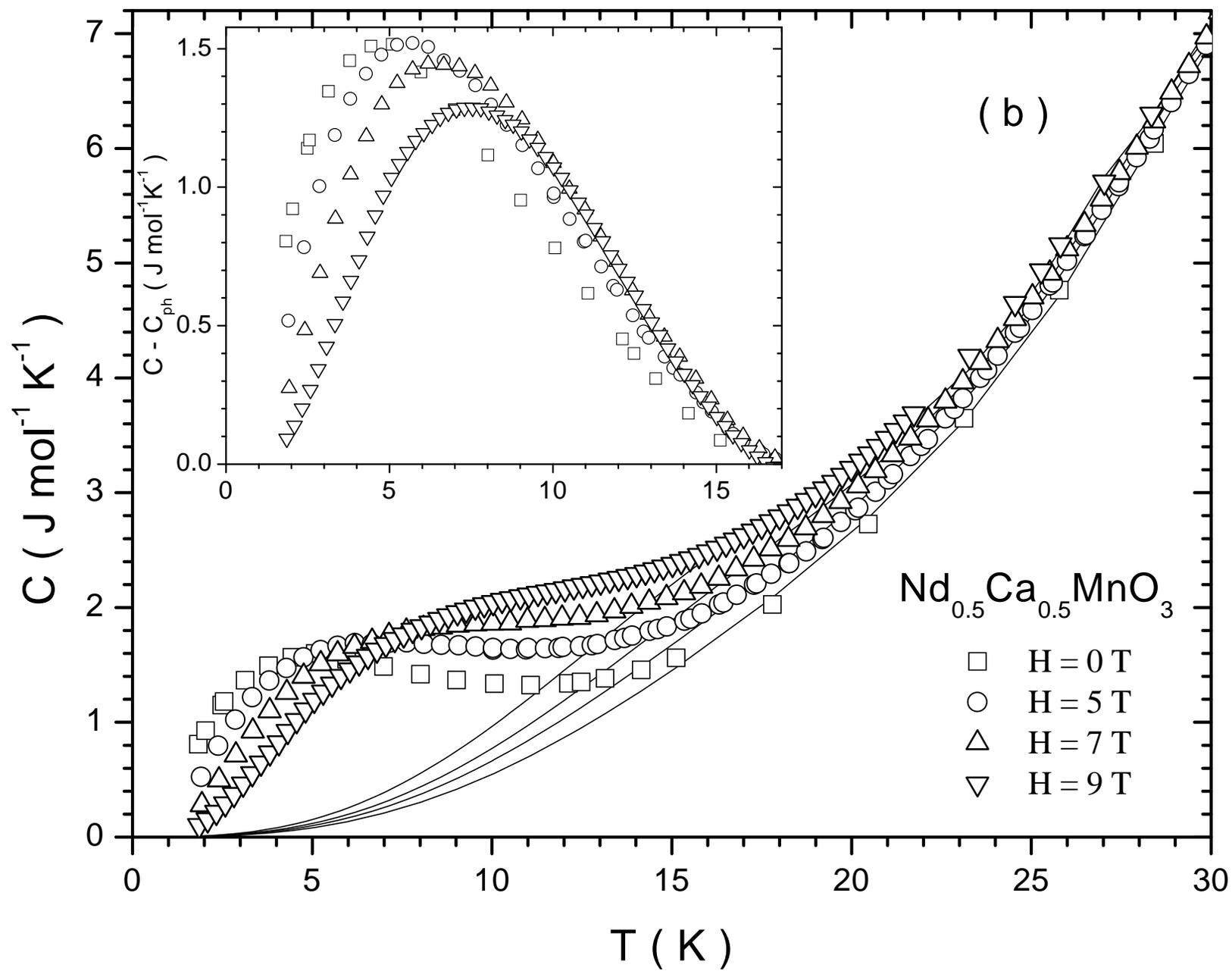
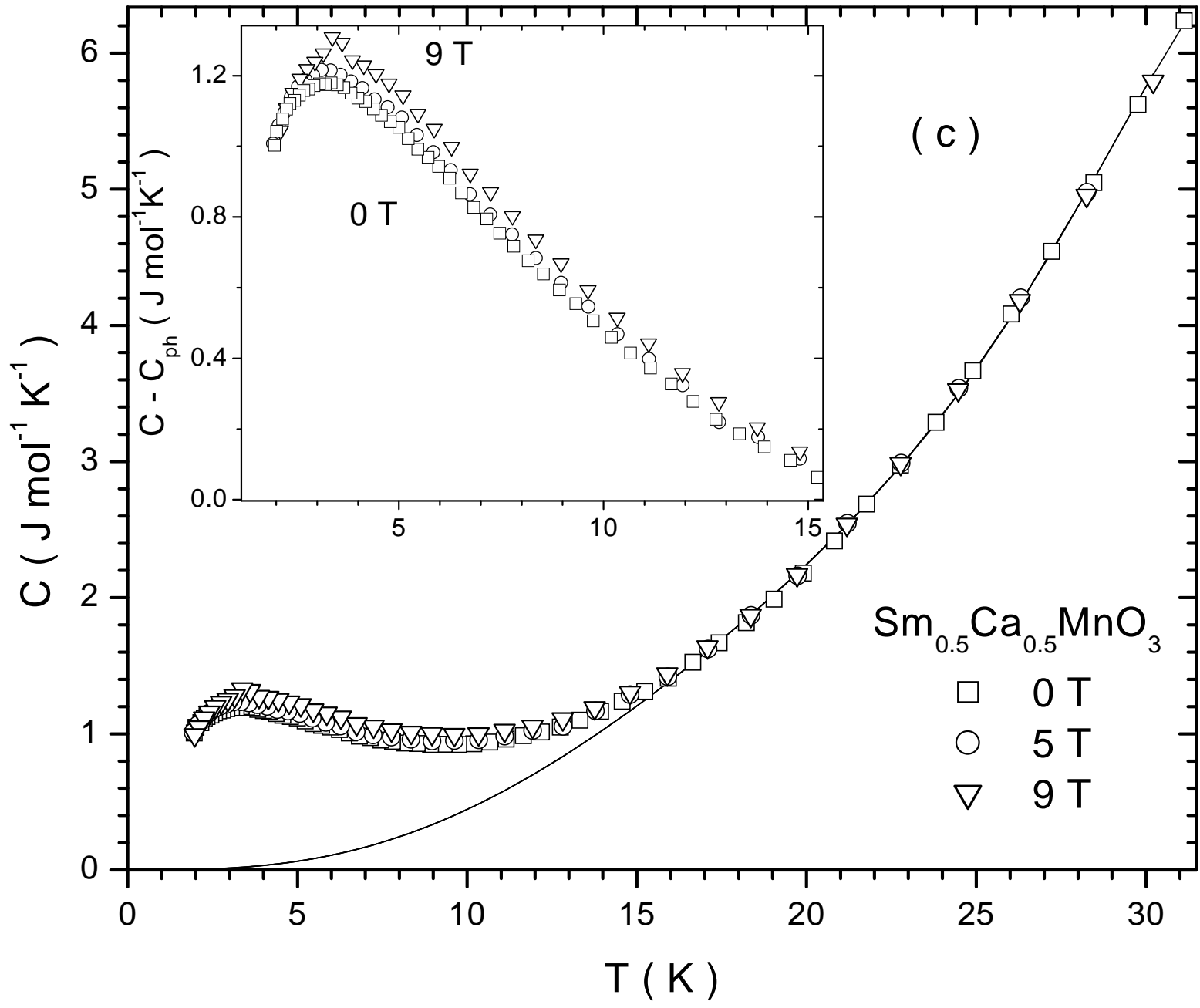
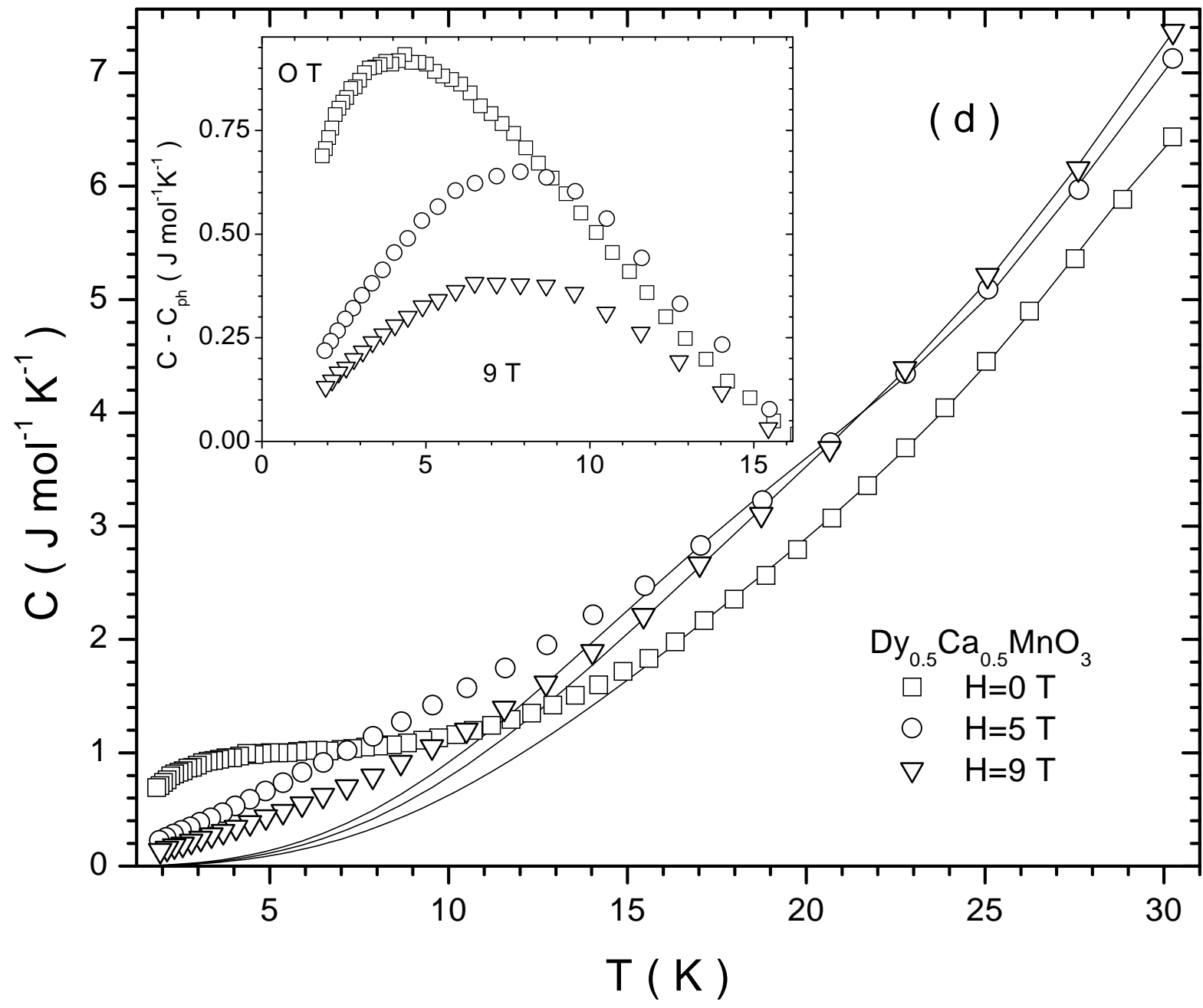
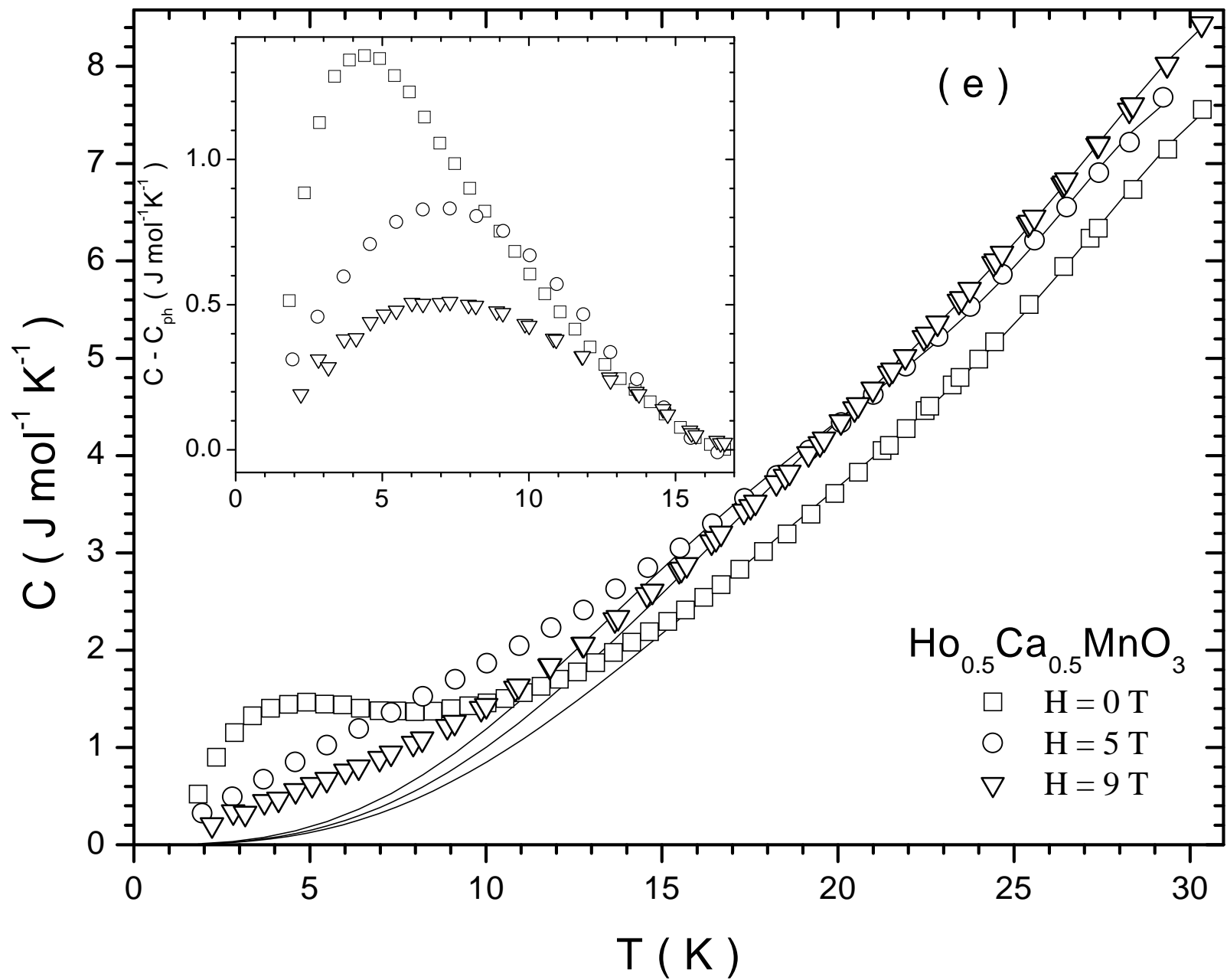
\end{document}